\documentclass[aps,pra,twocolumn,superscriptaddress,10pt,showpacs]{revtex4-1}
\pdfoutput=1 

\usepackage{graphicx}
\usepackage{subcaption}
\usepackage{amsmath}
\usepackage{gensymb}
\usepackage{epstopdf}

\usepackage{etoolbox}
\apptocmd{\sloppy}{\hbadness 10000\relax}{}{}

\begin{document}

%\preprint{}

\title{Generation of light with multimode time-delayed entanglement using storage in a solid state spin-wave quantum memory}


\author{Kate R. Ferguson}
%\email[Corresponding author \n]{katherine.ferguson@anu.edu.au}
\affiliation{Centre for Quantum Computation and Communication Technology, Laser Physics Centre, Australian National University, Canberra, Australian Capital Territory, 2601, Australia}
\author{Sarah E. Beavan}
\affiliation{European Space Agency, NL-2200 AG, Noordwijk, The Netherlands}
\author{Jevon J. Longdell}
\affiliation{Jack Dodd Centre for Photonics and Ultra-Cold Atoms, Department of Physics, University of Otago, Dunedin, New Zealand}
\author{Matthew J. Sellars}
%\thanks{}
\affiliation{Centre for Quantum Computation and Communication Technology, Laser Physics Centre, Australian National University, Canberra, Australian Capital Territory, 2601, Australia}


\date{\today}

\begin{abstract}
Here we demonstrate generating and storing entanglement in a solid state spin-wave quantum memory with on-demand read out using the process of rephased amplified spontaneous emission (RASE). Amplified spontaneous emission (ASE), resulting from an inverted ensemble of Pr$^{3+}$ ions doped into a Y$_2$SiO$_5$ crystal, generates entanglement between collective states of the praseodymium ensemble and the output light. The ensemble is then rephased using a four-level photon echo technique. Entanglement between the ASE and its echo is confirmed and the inseparability violation preserved when the RASE is stored as a spin-wave for up to 5 $\mu$s. RASE is shown to be temporally multimode with almost perfect distinguishability between two temporal modes demonstrated. These results pave the way for the use of multimode solid state quantum memories in scalable quantum networks. \par

\end{abstract}

%\keywords{}

\maketitle

%\section{}
%\subsection{}
%\subsubsection{}

Photonic quantum memories are essential devices in quantum information science. The role of a quantum memory, which can store and retrieve information encoded on photons, is to enable the synchronization of probabilistic quantum processes e.g. in quantum communication \cite{Sangouard2011} and computing \cite{Knill2001}. Storing the information carried by photons requires strong interactions between single photons and matter. While sufficiently strong interactions can be obtained by placing individual quantum systems in high finesse cavities \cite{Specht2011}, atomic ensembles provide an attractive alternative. \par

Rare-earth ion-doped crystals (REIC) are particularly promising due to long coherence times on both the optical and hyperfine transitions \cite{Zhong2015} while being free from decoherence due to atomic motion and offering a platform for integration \cite{Marzban2015}. Currently the low bandwidth of REIC memories makes it difficult to interface them with spontaneous parametric down conversation (SPDC) sources. Many demonstrations have shown storage of entangled \cite{Clausen2011,Saglamyurek2011} and heralded single photons \cite{Rielander2014}. However, these non-classical states were stored in memories with fixed storage times and the ability to read stored states out on-demand is essential for the synchronizing functionality of a quantum memory. The difficulty of interfacing entangled pair sources and REIC quantum memories means that on-demand solid state quantum memory demonstrations have stored weak coherent states \cite{Hedges2010,Gundogan2015,Afzelius2016}. \par

The difficulty with interfacing sources also limits atomic gas quantum memories, however a solution was proposed by Duan, Lukin, Cirac and Zoller in 2001. In the DLCZ protocol, non-classical states are generated from atomic ensembles using Raman transitions that lead to non-classical correlations between atomic excitations and emitted photons \cite{Duan2001}. The collective atomic excitations, called spin waves, can be efficiently read out in a well-defined spatial mode due to constructive interference of the atoms involved. To date no quantum solid-state DLCZ implementation has been demonstrated \cite{Goldschmidt2013}. \par

In 2010 the rephased amplified spontaneous emission scheme (RASE) was proposed, which has strong parallels to DLCZ, generating a collective atomic state via the measurement of spontaneous emission and rephasing it using photon echo techniques\cite{Ledingham2010}. Using resonant photon echo techniques has the advantage that the bright driving fields are off when the signals are being detected, as opposed to the non-resonant Raman techniques used in DLCZ, increasing the possibility of low noise operation. Long optical coherence times are required to implement this technique, which are present in REICs but not in atomic gas systems. Spin-wave storage is inherent in the four-level version of RASE (4L-RASE) \cite{Beavan2012}. The advantage of the RASE and DLCZ schemes is that non-classical states are both generated and stored in the same protocol, so the entangled light is automatically the correct wavelength and bandwidth. This will greatly improve the ease of integration. \par

In the basic RASE scheme the population of an inhomogeneously broadened ensemble of "two-level" atoms is inverted. The gain of the ensemble then amplifies the input vacuum fluctuations, emitting amplified spontaneous emission (ASE). While ASE on its own is considered a noisy field, this noise is due to entanglement generated between it and the collective modes of the amplifying medium. The ensemble inhomogeneity ensures that the collective atomic state dephases, however for systems like REICs with long coherence times these internal degrees of freedom can be rephased in a manner analogous to a photon echo \cite{Kurnit1964}. This rephased amplified spontaneous emission (RASE) allows for the state of the amplifying medium to be read out as a second optical field, entangled with the ASE. \par

Experimental investigations of RASE have been previously investigated using both two levels \cite{Ledingham2012} and four levels \cite{Beavan2012}. The four level work used storage in a spin-wave but was only able to show classical correlations due to the challenging spectral filtering requirements associated with discrete-variable detection. The two level RASE experiment came closer to showing entanglement but didn't incorporate spin-wave storage. Here, for the first time we show entanglement between the time-separated ASE and RASE fields after storage of the RASE field as a spin-wave. Continuous-variable detection is used to spectrally discriminate emission to nearby hyperfine levels. \par

\begin{figure}[ht!]
\centering
\includegraphics[width=0.45\textwidth]{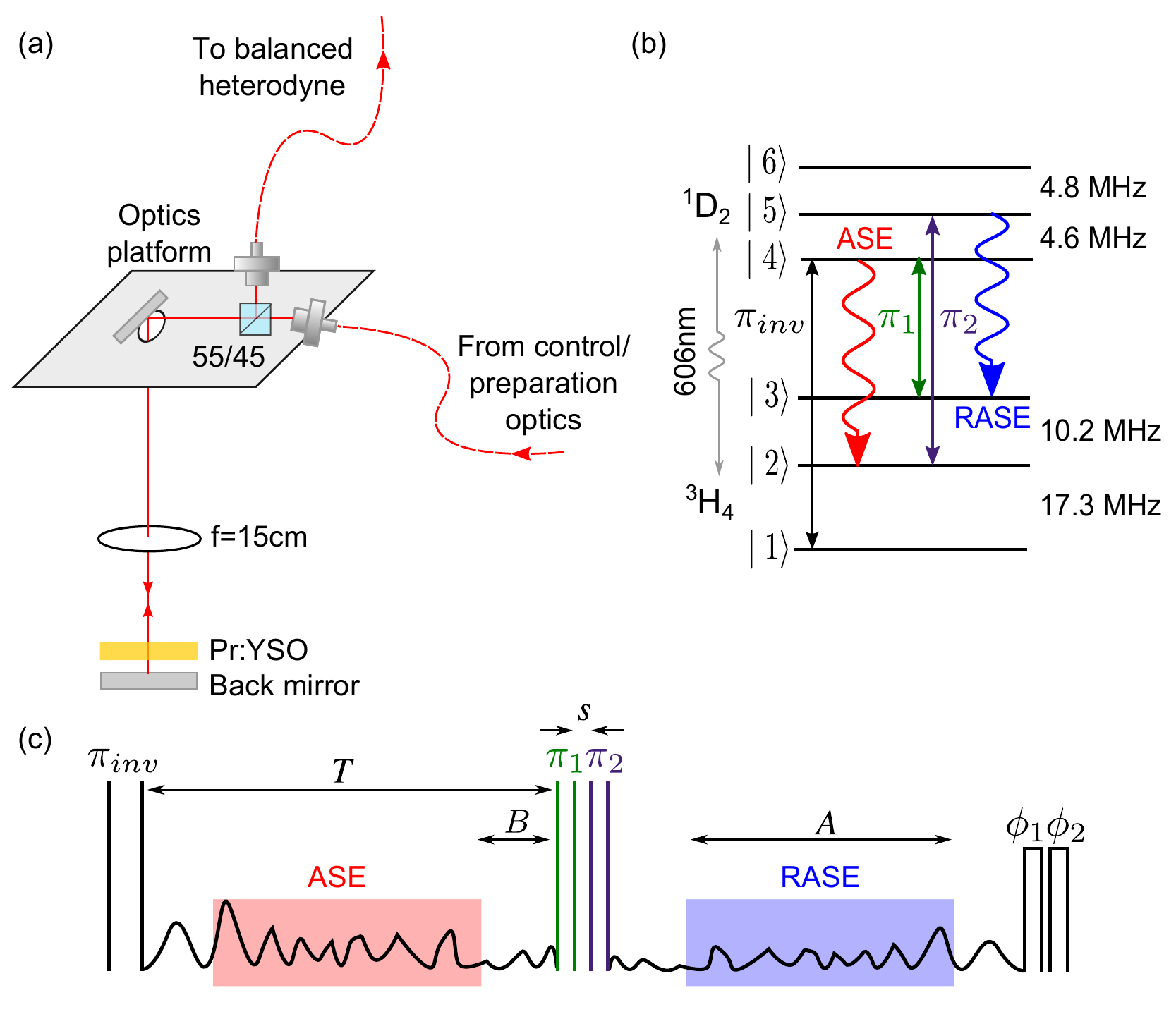}
\caption{\label{fig:exptSetup}(color online) (a) RASE setup. The 0.005\% Pr:YSO crystal, lens and back mirror are located at the bottom of a liquid helium bucket cryostat at 4 K. The cryostat only has vertical optical access so a top optics platform has fibre couplers steering the control and preparation beams into the cryostat and the reflected signal beam to the balanced heterodyne detection system. (b) Hyperfine levels of the ground $^3H_4$ and excited $^1D_2$ manifold in Pr:YSO. For the RASE experiment a sub-ensemble of ions is selected using spectral holeburning and prepared to be initially in state $|1\rangle$. The frequencies used to apply the 4L-RASE protocol are marked. (c) Pulse sequence used in a single shot of the 4L-RASE experiment. After the inversion pulse $\pi_{inv}$ is applied, the ensemble is allowed to spontaneously emit for $T$ before the rephasing pulses $\pi_1$ = 1.5 $\mu$s and $\pi_2$ = 2.2 $\mu$s are applied. These are separated by a time $s$, the storage time on the spin states. $A$ is the length of the ASE and RASE windows used for calculating the quadrature values. $B$ is the delay between the end (start) of the ASE (RASE) window and the rephasing pulses to allow the detectors to recover from saturation due to the intense $\pi$-pulses. After the sequence two weak phase reference pulses, $\phi_1$ and $\phi_2$, are applied to correct for the frequency dependent phase offsets shot to shot.}
\end{figure}

The 4L-RASE pulse sequence is depicted in the frequency and temporal domain in Fig \ref{fig:exptSetup}(b) and \ref{fig:exptSetup}(c) respectively. The four-level sequence rephases coherence generated between two levels (here $|2\rangle \leftrightarrow|4\rangle$) while transferring it to two completely different levels ($|3\rangle \leftrightarrow |5\rangle$). This is achieved by applying two sequential rephasing $\pi$-pulses driving transitions $|3\rangle \leftrightarrow|4\rangle$ and $|2\rangle \leftrightarrow |5\rangle$. The free-induction decay (FID) resulting from the two $\pi$-pulses is now spectrally resolvable from both the ASE and RASE emission. Following the application of the first $\pi$-pulse the entangled state is stored on the long-lived hyperfine ground states as a spin-wave before being transferred back to the optical transition by the second $\pi$-pulse. \par

The sample used in this experiment was a 2 x 4 x 5 mm 0.005\% Pr$^{3+}$ : Y$_2$SiO$_5$ (Pr:YSO) crystal cooled to 4.2 K using exchange gas cooling in a liquid helium bucket cryostat. There is only vertical optical access to the cryostat, so a back mirror reflects the light back through the sample to an optics platform on top of the cryostat. The experimental arrangement is shown in Fig. \ref{fig:exptSetup}(a). The energy level structure of Pr:YSO is depicted in Fig. \ref{fig:exptSetup}(b). The lifetime for the excited state is $T_1$ = 164 $\mu$s \cite{Equall1995} and the dephasing time was measured here to be $T_2^{\ast}$ 151 $\mu$s . The lifetime between the ground state hyperfine levels is $\sim$ 200 s \cite{Holliday1993} and the coherence time is $\sim$ 500 $\mu$s at zero magnetic field \cite{FravalE}, so potentially long storage times are obtainable. All the optical transitions are weakly allowed with the transition strengths reported in Ref \cite{Nilsson2004}. \par

The optical decay time for the four-level sequence used for the RASE scheme was measured to be 53 $\mu$s. This is roughly a third of the coherence time for the standard two-level echo. The discrepancy is attributed to a component of inhomogeneity on the spin levels that is not rephased using the four-level sequence \cite{Beavan2011}.\par

The optical $^3H_4 \rightarrow ^1D_2$ transition is inhomogeneously broadened to 3 GHz, two orders of magnitude larger than the $\sim$ 10 MHz hyperfine splittings. Before each shot of the RASE experiment, a subgroup of ions with 150 kHz inhomogeneity is selected and the population initialized into $|1\rangle$ using spectral hole-burning techniques similar to those described in Ref. \cite{Nilsson2004,Beavan2012}. \par

After selecting out the ensemble a single shot of the RASE experiment is performed. The ensemble preparation takes $\sim$ 100 ms limiting the repetition rate of the experiment to 10 Hz. The temporal sequence is outlined in Fig. \ref{fig:exptSetup}(c). For the first experiment $s$ = 0 $\mu$s, $T$ = 18.5 $\mu$s, $A$ = 10 $\mu$s and $B$ = 5 $\mu$s. The inversion pulse creates a gain feature with optical depth of $\alpha l$ = 2.35 \footnote{See Supplementary Material at URL for details of gain measurements, phase correction, the windowing of the ASE and RASE signals and the inseparability criterion model.}. The control beam is gated on with a double-pass acousto-optic modulator (AOM). Shot noise limited heterodyne detection is used to characterize the ASE and RASE fields by measuring the variances of the amplitude $\hat{x}$ and phase $\hat{p}$ quadratures of light. The phase of the interferometer was not locked and was different shot to shot. The phase between the different frequency RF pulses was not constant either. Two phase reference pulses applied after the 4L-RASE sequence allowed the data to be corrected for these two issues \cite{Note1}. \par

The first way of characterizing the correlation between the ASE and RASE fields is by evaluating the cross-correlation of the two fields. The rephasing sequence ideally results in a time-reversed, conjugated version of the ASE field so the cross-correlation is defined as

\begin{equation}
\label{eq:crossCor}
C(\tau) = \int A(t)R(\tau-t)dt
\end{equation}

where $A(t)$ ($R(t)$) is the amplitude of ASE (RASE) field and $\tau$ = 0 has been chosen to correspond to the centre of the two rephasing pulses. The integral is over all time and $A(t)$ and $R(t)$ are windowed such that they are zero outside the time windows shown in Fig. \ref{fig:exptSetup}(c). \par

\begin{figure}[ht!]
\centering
\includegraphics[width=0.45\textwidth]{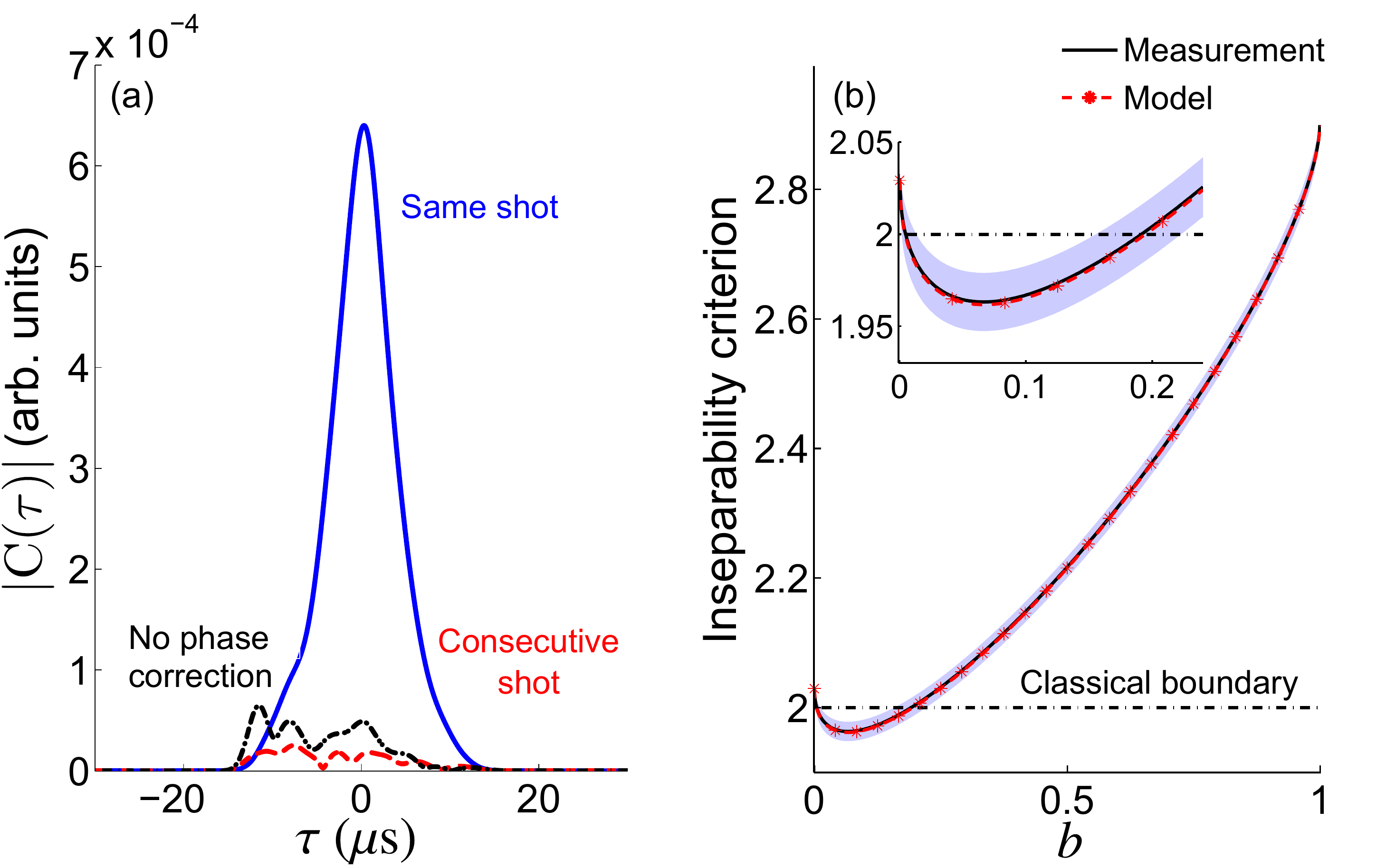}
\caption{\label{fig:insepViolation}(color online) (a) The cross-correlation function between the ASE and RASE fields [Eq. \ref{eq:crossCor}] of the same shot, consecutive shots and when the phase correction is not applied averaged over 8000 trials. (b) The inseparability criterion [Eq. \ref{eq:insep}] as a function of $b$ when $s$ = 0. The ASE(RASE) variance is 1.453 $\pm$ 0.023 (1.015 $\pm$ 0.016) normalized to the vacuum. The shaded area indicates 1$\sigma$ confidence. Inset shows a close up of the minimum of the inseparability criterion. }

\end{figure}

Figure \ref{fig:insepViolation}(a) shows the mean cross-correlation between the ASE and RASE of the same shots, of different shots and when the phase correction is not applied. There is a distinct correlation peak only present when comparing the ASE and RASE of the same shot, so there is no evidence of correlations due to coherent effects. This confirms a time-separated correlation between the ASE and RASE fields. The 7 $\mu$s temporal width of the correlation peak corresponds to a 65 kHz ASE bandwidth. \par

To test the quantum nature of this time-separated correlation the inseparability criterion for continuous variable states created by Duan \textit{et al.} is used \cite{Duan2000}. A maximally entangled state can be expressed as a co-eigenstate of a pair of EPR-type operators

\begin{equation} \label{eq:eprOp}
\hat{u} = \sqrt{b}\hat{x}_1 + \sqrt{1-b}\hat{x}_2,  \;\;\;\;\;\;\;\;\;\; \hat{v} = \sqrt{b}\hat{p}_1 - \sqrt{1-b}\hat{p}_2,
\end{equation}

where $b \in [0,1]$ is a weighting parameter describing the weight given to the ASE and RASE fields. The subscript 1(2) indicates the ASE(RASE) field. \par

For any separable state, the total variance of $\hat{u}$ and $\hat{v}$ satisfies

\begin{equation}
\label{eq:insep}
\langle (\Delta \hat{u})^2\rangle + \langle (\Delta \hat{v})^2\rangle \geq 2.
\end{equation}

For inseparable states, the total variance is bound from below by zero. \par

Heterodyne detection provides simultaneous, noisy measurements of both the light quadratures \cite{Yuen1983} as opposed to homodyne detection which provides a good measure of just one quadrature. The same inseparability criterion can be used for both detection methods however the the size of the correlation will be reduced by a factor of 2 when using a heterodyne detector \cite{Ledingham2012}. \par

The complex valued heterodyne signal was windowed with 10 $\mu$s temporal functions that are a convolution of a top hat function and a Gaussian and then integrated to obtain values of $\hat{x}$ and $\hat{p}$ for both the ASE and RASE time windows. The spectral width of this window was optimized to match the spectral profile of the signals, giving the strongest correlation \cite{Note1}. \par

When $b$ = 1(0) the variance only consists of the ASE (RASE) field summed over both quadratures. The uncorrelated case is a straight line between these two values. Fig. \ref{fig:insepViolation}(b) shows the inseparability criterion verses $b$. There is a clear dip indicating a correlation between the ASE and RASE fields. At the lowest point $\langle (\Delta \hat{u})^2\rangle + \langle (\Delta \hat{v})^2\rangle$ = 1.964 at $b$ = 0.068 violating the inseparability criterion with 98.6\% confidence, a 2.2$\sigma$ violation. The error in the variance was calculated as in \cite{Ledingham2012}. The low value of $b$ is due to the difference in size of the ASE and RASE fields due to the low rephasing efficiency of 3.2\%, determined by the ratio of the signal variances. \par

Expected values for the inseparability criterion are calculated based on a simple model using the ASE variance and rephasing efficiency. The model assumes the ASE and RASE fields are initally maximally entangled and this entanglement is then degraded by loss. The loss due to the two beamsplitters, reflection and the detector efficiencies are modelled as a single beamsplitter in both the ASE and RASE modes and the rephasing efficiency is modelled as an additional beamsplitter in the RASE mode \cite{Note1}. The measured inseparability tracks almost perfectly with the modelled criterion indicating very little noise is added during the rephasing process. \par 

The limitation in this experiment is the rephasing efficiency, which should increase with increasing optical depth (OD)\cite{Ledingham2012,Stevenson2014}, however it was seen to saturate at $\sim$ 3\%. Perfect rephasing $\pi$-pulses were not possible in the present configuration, where the control fields and signals were in the same mode, and the optically thick sample distorted the pulses to a larger degree as the OD increased. The distortion introduced additional noise such that the measured inseparability criterion no longer agreed with the model for large OD. The efficiency can theoretically approach 100\% by placing the sample in a low finesse cavity \cite{Williamson2014}. In this case, the trade off between rephasing pulse uniformity and optical depth could be avoided by applying the $\pi$-pulses off axis. Efficiency enhancement has similarly been shown for DLCZ by placing cold atoms in a ring cavity \cite{Bao2012}. \par

\begin{figure}[ht!]
\centering
\includegraphics[width=0.45\textwidth]{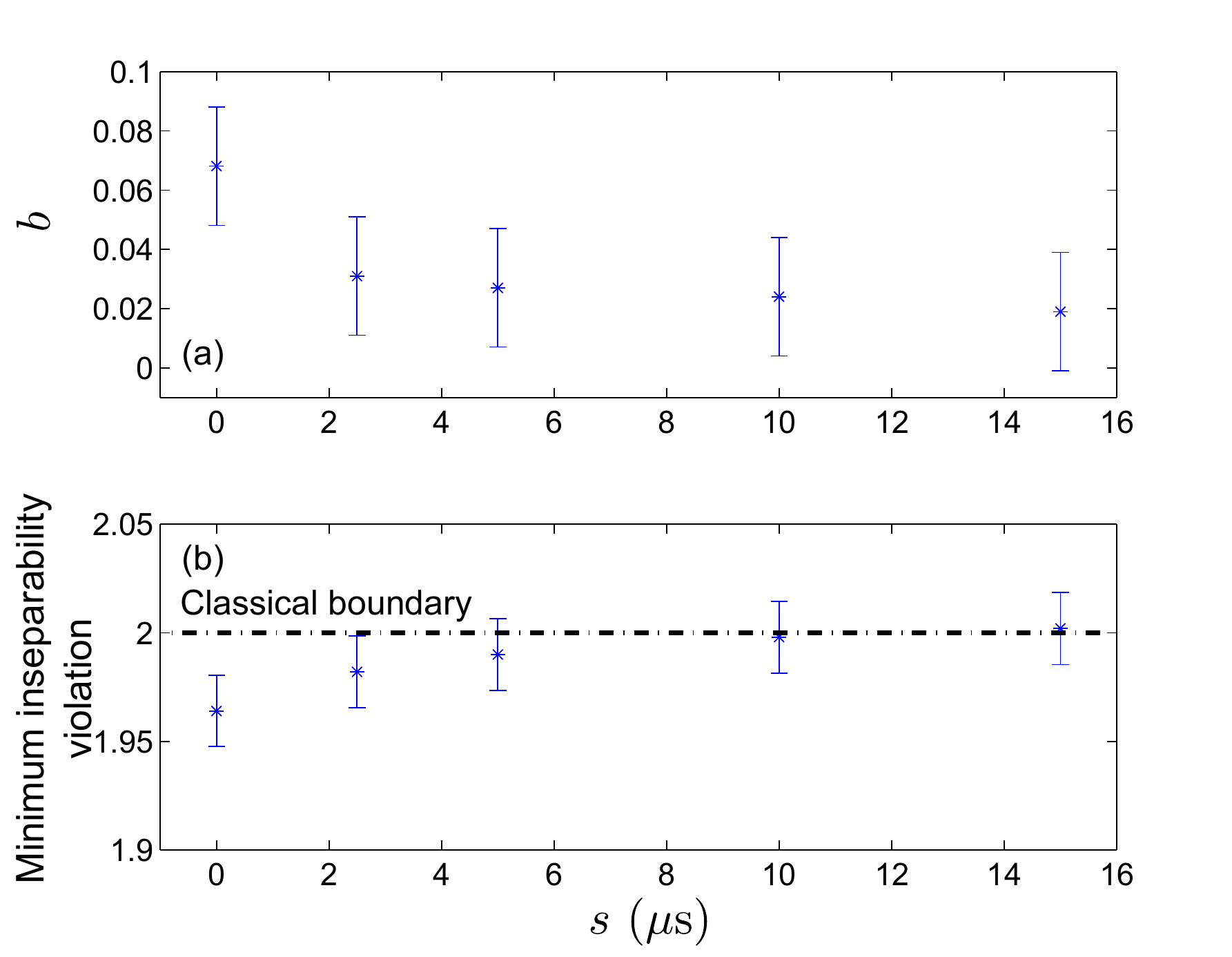}
\caption{\label{fig:delay}(color online) (a) $b$ and (b) the minimum inseparability violation with increasing spin-wave storage time $s$. All errorbars are 1$\sigma$.}
\end{figure}

Next we investigated the spin wave storage of the quantum memory by varying $s$. The total storage time is $T_s$ = $2B + s$. The dephasing caused by the inhomogeneous broadening on the spin states is not rephased by the four-level sequence resulting in a decay of $b$ with increasing $s$ (Fig. \ref{fig:delay}(a)) as the rephasing efficiency drops. In Fig. \ref{fig:delay}(b) the minimum inseparability violation gets correspondingly worse until the criterion is at the classical boundary for $s$ = 10 $\mu$s and $s$ = 15 $\mu$s ($T_s$ = 20-25 $\mu$s). It should be possible to increase the storage time by applying rephasing RF $\pi$-pulses to the hyperfine levels \cite{Fraval2004,Zhong2015}. \par

RASE is based on photon echo rephasing techniques, and as such should be inherently temporally multimode. In quantum repeater applications the addition of multimode memories able to store and retrieve $N$ different modes while preserving their distinguishability can increase the overall success rate of the repeater by that factor $N$ \cite{Simon2007}. \par

\begin{figure}[ht!]
\centering
\includegraphics[width=0.45\textwidth]{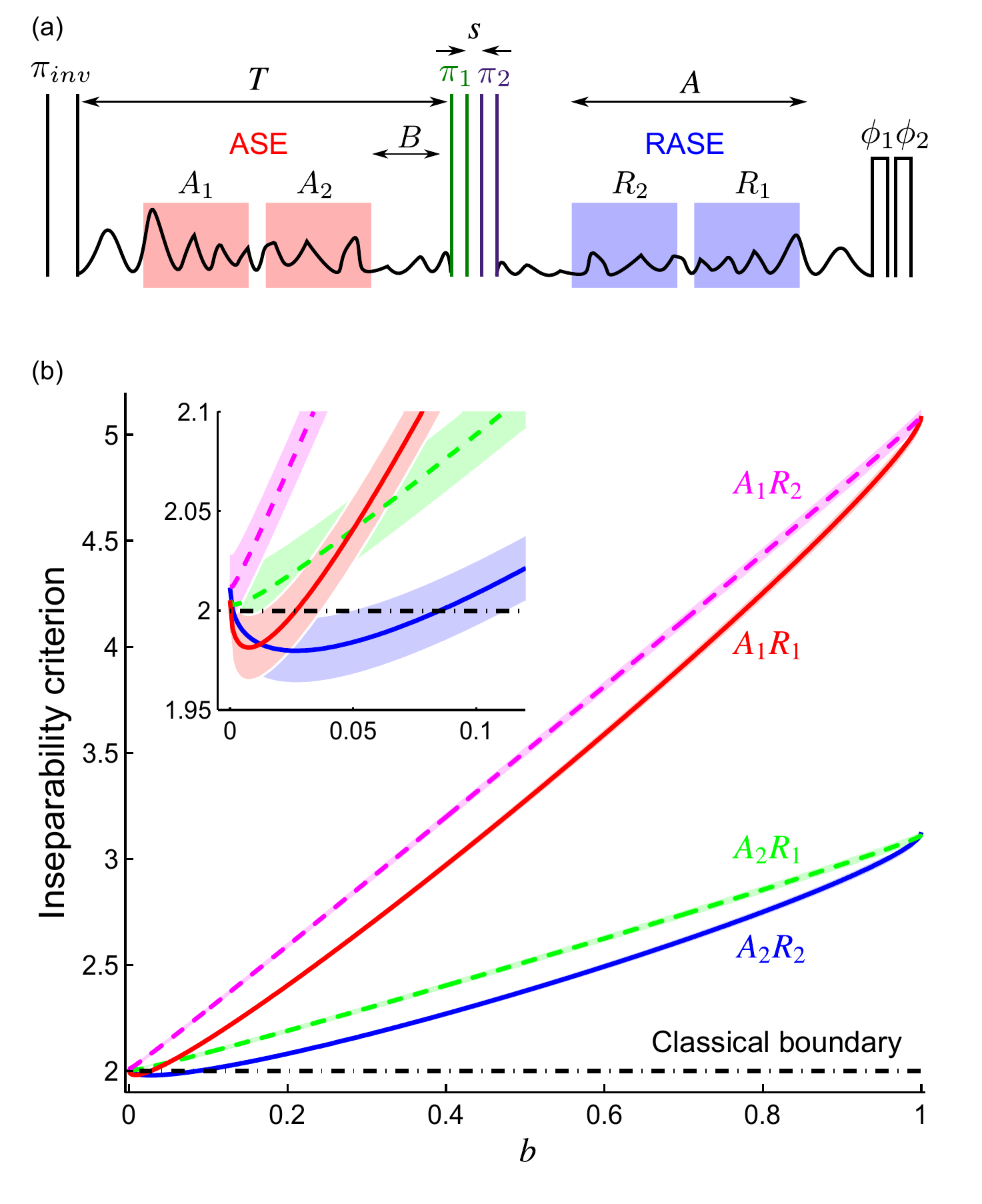}
\caption{\label{fig:multimode}(color online) (a) Pulse sequence used for the multimode 4L-RASE experiment showing the labelling of the two temporal modes. $A$ = 20 $\mu$s, $B$ = 5 $\mu$s, $T$ = 28.5 $\mu$s and $s$ = 0 $\mu$s. (b) Inseparability criterion for the four combinations of the two temporal modes. The solid lines are the time-symmetric combinations showing a correlation. The dashed lines are the non-symmetric cases and are uncorrelated. The shaded areas indicate 1$\sigma$ confidence. Inset shows a close up of the minimum of the criterion.}
\end{figure}

To test the multimode capability of RASE the signal windows in Fig. \ref{fig:exptSetup}(c) are increased to $A$ = 20 $\mu$s allowing two 10 $\mu$s temporal windows per shot. The labeling of the two temporal windows is shown in Fig. \ref{fig:multimode}(a). The inseparability criteria for the four possible combinations of windows are shown in Fig. \ref{fig:multimode}(b). The two combinations where the ASE and RASE windows are time-symmetric around the rephasing pulses have a correlation between the two fields while the non-symmetric combinations are almost completely uncorrelated, evidenced by the straight line criterion. The distinguishability between the different temporal modes is therefore almost perfectly preserved. \par

The two correlated temporal modes violate the inseparability criterion with 86\% and 89\% confidence for $A1R1$ and $A2R2$ respectively. $b$ is $\sim$ 3.5x smaller for $A1R1$ corresponding to a reduced rephasing efficiency further from the $\pi$-pulses. \par

In conclusion, we have demonstrated a means of generating entanglement and storing it on the spin states of a Pr:YSO crystal through rephasing spontaneous emission. We confirmed the temporal multimode capability of RASE inherent in the rephasing process. Combining temporal multiplexing and storage on the spin states allows the possibility of simultaneous read of different temporal modes from neighbouring memories to perform entanglement swapping operations.  While the current demonstration used continuous-variable detection, this scheme is equally capable of generating on-demand single photons. RASE takes advantage of the unique coherence properties of REICs to generate and store entangled states in a single protocol, paving the way to making scalable, solid state quantum information processing architectures. \par

\appendix
\section{Gain measurement and expected signal variances}

The gain of the inverted feature was measured by applying a long, weak probe windowed over the same region as the ASE signal window. The probe had 100 kHz bandwidth and was stepped in 20 kHz steps from the centre frequency of the inversion to map out the gain profile. The ratio of the Fourier transform of the probe with and without the inversion pulse applied gave the gain $G = \textrm{ln}(\alpha l)$ where $\alpha l$ is the optical depth. From fitting the gain profile, we determined $\alpha l$ = 2.35 and the bandwidth = 160 kHz. \par

The expected signal variance can be calculated from the optical depth using the model developed by Ledingham \textit{et. al} \cite{Ledingham2012}. To accurately determine the variance the loss due to the beamsplitters in the experiment must be considered. The loss will reduce the variances as shown in the discussion on modelling the inseparability criterion. The predicted ASE variance is 5.74 for the measured $\alpha l$ = 2.35. The observed ASE variance was 1.453 $\pm$ 0.023, roughly 10\% of the predicted value. \par

A possible explanation for the gain discrepancy arises from the optics geometry inside the cryostat shown in Fig. 1(a) of the text. If the waist of the beam is not perfectly on the mirror behind the crystal, then there will be a spatial mismatch between the input and reflected beams. When the gain is measured by applying a probe the detected light has necessarily traveled through the crystal twice, giving a total interaction length of 4 mm. During the emission of ASE, the input vacuum state that will couple into the detection mode will only be partially amplified on the initial pass through the crystal. The reflected state will be fully amplified by the gain region in the crystal. The total gain the vacuum state sees will be between the single pass ($\alpha l$ = 1.175) and double pass ($\alpha l$ = 2.35) gain of the ensemble. The measured variance falls within this range. The spatial mode mismatch between the ASE mode and the input weak coherent state used to probe the gain is also the probable explanation for the discrepancy between the measure gain bandwidth (160 kHz) and the bandwidth of the ASE determined from the cross-correlation (65 kHz) in Fig. 2(a) of the text. \par

\section{Phase correction}

Heterodyne detection is sensitive to the phase difference between the signal and the local oscillator (LO) beams. Mechanical vibrations and thermal fluctuations in the lab caused the phase relationship between the two beams to vary randomly between different shots of the experiment. Therefore, to compare multiple shots of the RASE experiment, and build up the statistics required to show entanglement, it was necessary to correct the phase of each shot. \par

The oscillators clocking the RF drivers for the control and LO AOMs, and the digital oscilloscope were synchronised ensuring a constant frequency offset. The phase variation over a single 200 $\mu$s shot of the experiment was measured to be, at worst, on the 1$\degree$ level. In addition, the phase noise on the frequency stabilised Coherent 699 dye laser over the timescale of the experiment was also less than 1$\degree$. \par 

Two factors had to be corrected between different shots. First, the global phase change of the interferometer. The second factor was a timing jitter in the triggering of the oscilloscope. This was present because the clock for the RF sources was a factor of three larger than the clock for the oscilloscope. As a result, the oscilloscope could trigger at one of three different points in the RF clock cycle. The timing jitter was measured to be approximately 3 ns and introduced a frequency dependent phase shift. As a result, two phase reference pulses at different frequencies were required to correct the phase between shots of the experiment. The first pulse was used to correct for the global interferometer phase and the second pulse was then used to calculate the small frequency dependent phase shift due to the timing jitter. After correction the phase of the different frequency signals was typically around 3$\degree$. \par

\section{Windowing the signals}

The time regions for the ASE and RASE, indicated by colored boxes in Fig. 1(c) of the text, were temporally windowed with a convolution of a top hat function, length $h$, and a Gaussian, width $w$, which were varied to optimize the correlation between the two fields. The correlation will be optimized when this windowing function most closely matches the temporal profile of the ASE/RASE. For the inseparability criterion shown in Fig. 2(b) of the text, the windowing function was optimized with $h$ = 7 $\mu$s and $w$ = 600 kHz. The edges of the window were cut off at 10 $\mu$s, the length of the signal data collected, to ensure no signal too close to the $\pi$-pulses was included. \par

After windowing, each shot was then digitally filtered with a 500 kHz Gaussian filter and beat to DC. The vacuum had to be separately filtered at both the ASE and RASE frequency to correctly normalize the different frequency signals. \par

\section{\label{insepModel} Modelled inseparability criterion}

Here we describe how experimental losses and the rephasing efficiency are included in the model of the inseparability criterion \cite{Duan2000} used in Fig. 2 in the paper. From Eq. 2 in the paper, the left side of Eq. 3 can be written as

\begin{align} \label{eq:insepExpand}
\textrm{var}(\hat{u}) + \textrm{var}(\hat{v}) &= b\langle\hat{x}_1^2\rangle + (1-b)\langle\hat{x}_2^2\rangle \notag \\ 
&+ 2\sqrt{b(1-b)}\langle\hat{x}_1\hat{x}_2\rangle \notag \\ 
&+ b\langle\hat{p}_1^2\rangle + (1-b)\langle\hat{p}_2^2\rangle \notag \\ 
&- 2\sqrt{b(1-b)}\langle\hat{p}_1\hat{p}_2\rangle.
\end{align}

We used heterodyne detection rather than homodyne detection for these experiments mostly because we were dealing with signals at several different frequencies. Heterodyne detection makes a 'noisy' measurement of both quadratures of light simultaneously \cite{Yuen1983}. The measured quadratures values appear to have passed through a 50:50 beamsplitter with the vacuum entering the unused port. \par

There is additional loss due to the 55:45 beamsplitter on top of the cryostat platform in Fig. 1 in the paper. The reflection losses and detector efficiencies along with the losses due to the two beamsplitters can be modelled as a single beamsplitter with variable transmission $\frac{1}{L}$. The measured quadrature value $\hat{x}_i$ after this loss is

\begin{equation}
\hat{x}_i \rightarrow \sqrt{\frac{1}{L}}\hat{x}_i + \sqrt{\bigg(1-\frac{1}{L}\bigg)}\hat{v}_i,
\end{equation}  

where $i$ = 1 or 2 and $\hat{v}_i$ is the vacuum state that enters the unused port of the beamsplitter. \par

The rephasing inefficiency can be introduced as a variable beamsplitter with transmission $\frac{1}{e}$ that only affects the RASE quadratures. \par

Substituting these two sources of loss into Eq. \ref{eq:insepExpand} and remembering the variance of the vacuum is 1 and the covariance between $\hat{x}_i$ and the vacuum is zero we get

\begin{align} \label{eq:insepLoss}
\langle \hat{u}^2\rangle &= b\bigg(\frac{1}{L}\langle{x}_1^2\rangle + \bigg(1-\frac{1}{L}\bigg)\bigg) \notag \\
&+ (1-b)\bigg(\frac{1}{L}\bigg(\frac{1}{e}\langle\hat{x}_2^2\rangle + \bigg(1-\frac{1}{e}\bigg)\bigg) + \bigg(1-\frac{1}{L}\bigg)\bigg) \notag \\
&+ 2\sqrt{b(1-b)}\frac{1}{L}\sqrt{\frac{1}{e}}\langle\hat{x}_1\hat{x}_2\rangle,
\end{align}

with a similar expression for $\langle \hat{v}^2\rangle$. \par 

When comparing Eq. \ref{eq:insepExpand} and Eq. \ref{eq:insepLoss}, the covariance term is reduced by a factor of $\frac{1}{L}\sqrt{\frac{1}{e}}$, hence the correlation strength between the ASE and RASE fields is also decreased by this factor. The covariance is less dependent on the rephasing efficiency as it only affects the RASE component. \par

In this case of perfect rephasing and no loss, the ASE and RASE fields will be maximally entangled. In this case, the inseparability criterion dips to 0 for $b$ = 0.5, i.e. when equal weight is given to the two fields. For this experiment, the loss was determined to be $\frac{1}{L} = \frac{1}{4}$ and the rephasing efficiency measured as $\frac{1}{e} = 0.032$ by taking the ratio of the ASE and RASE variances. The loss reduces the possible inseparability violation by a factor of 4, resulting in a lowest possible dip to 1.5 at $b$ = 0.5. When the imperfect rephasing is also included the minimum inseparability no longer occurs when equal weight is given to the ASE and RASE fields. For the measured rephasing efficiency in this experiment, there is a lowest possible inseparability dip to 1.962 occurring at $b$ = 0.068. \par

By setting the ASE and RASE quadratures to be vacuum states we can determine that while the size of the inseparability violation will be reduced by the loss and inefficiency, the threshold for inseparability does not change. \par

The measured ASE quadratures were used to calculate the ASE (and RASE assuming the fields are initially maximally entangled) quadratures before loss, which were then input, along with the measured rephasing efficiency, into the model. \par

\begin{acknowledgments}
This work was supported by the Australian Research Council Centre of Excellence for Quantum Computation and Communication Technology (Grant No. CE110001027). M.J.S. was supported by an Australian Research Council Future Fellowship (Grant No. FT110100919). J.J.L. would like to acknowledge the Marsden Fund (Contract No. UOO1520) of the Royal Society of New Zealand. The authors would like to acknowledge the support of the Australian Defense Science and Technology Group. 
\end{acknowledgments}

\bibliography{arXivPaper}

\begin{thebibliography}{30}%
\makeatletter
\providecommand \@ifxundefined [1]{%
 \@ifx{#1\undefined}
}%
\providecommand \@ifnum [1]{%
 \ifnum #1\expandafter \@firstoftwo
 \else \expandafter \@secondoftwo
 \fi
}%
\providecommand \@ifx [1]{%
 \ifx #1\expandafter \@firstoftwo
 \else \expandafter \@secondoftwo
 \fi
}%
\providecommand \natexlab [1]{#1}%
\providecommand \enquote  [1]{``#1''}%
\providecommand \bibnamefont  [1]{#1}%
\providecommand \bibfnamefont [1]{#1}%
\providecommand \citenamefont [1]{#1}%
\providecommand \href@noop [0]{\@secondoftwo}%
\providecommand \href [0]{\begingroup \@sanitize@url \@href}%
\providecommand \@href[1]{\@@startlink{#1}\@@href}%
\providecommand \@@href[1]{\endgroup#1\@@endlink}%
\providecommand \@sanitize@url [0]{\catcode `\\12\catcode `\$12\catcode
  `\&12\catcode `\#12\catcode `\^12\catcode `\_12\catcode `\%12\relax}%
\providecommand \@@startlink[1]{}%
\providecommand \@@endlink[0]{}%
\providecommand \url  [0]{\begingroup\@sanitize@url \@url }%
\providecommand \@url [1]{\endgroup\@href {#1}{\urlprefix }}%
\providecommand \urlprefix  [0]{URL }%
\providecommand \Eprint [0]{\href }%
\providecommand \doibase [0]{http://dx.doi.org/}%
\providecommand \selectlanguage [0]{\@gobble}%
\providecommand \bibinfo  [0]{\@secondoftwo}%
\providecommand \bibfield  [0]{\@secondoftwo}%
\providecommand \translation [1]{[#1]}%
\providecommand \BibitemOpen [0]{}%
\providecommand \bibitemStop [0]{}%
\providecommand \bibitemNoStop [0]{.\EOS\space}%
\providecommand \EOS [0]{\spacefactor3000\relax}%
\providecommand \BibitemShut  [1]{\csname bibitem#1\endcsname}%
\let\auto@bib@innerbib\@empty
%</preamble>
\bibitem [{\citenamefont {Sangouard}\ \emph {et~al.}(2011)\citenamefont
  {Sangouard}, \citenamefont {Simon}, \citenamefont {de~Riedmatten},\ and\
  \citenamefont {Gisin}}]{Sangouard2011}%
  \BibitemOpen
  \bibfield  {author} {\bibinfo {author} {\bibfnamefont {N.}~\bibnamefont
  {Sangouard}}, \bibinfo {author} {\bibfnamefont {C.}~\bibnamefont {Simon}},
  \bibinfo {author} {\bibfnamefont {H.}~\bibnamefont {de~Riedmatten}}, \ and\
  \bibinfo {author} {\bibfnamefont {N.}~\bibnamefont {Gisin}},\ }\href
  {\doibase 10.1103/RevModPhys.83.33} {\bibfield  {journal} {\bibinfo
  {journal} {Rev. Mod. Phys.}\ }\textbf {\bibinfo {volume} {83}},\ \bibinfo
  {pages} {33} (\bibinfo {year} {2011})}\BibitemShut {NoStop}%
\bibitem [{\citenamefont {Knill}\ \emph {et~al.}(2001)\citenamefont {Knill},
  \citenamefont {Laflamme},\ and\ \citenamefont {Milburn}}]{Knill2001}%
  \BibitemOpen
  \bibfield  {author} {\bibinfo {author} {\bibfnamefont {E.}~\bibnamefont
  {Knill}}, \bibinfo {author} {\bibfnamefont {R.}~\bibnamefont {Laflamme}}, \
  and\ \bibinfo {author} {\bibfnamefont {G.~J.}\ \bibnamefont {Milburn}},\
  }\href@noop {} {\bibfield  {journal} {\bibinfo  {journal} {Nature}\ }\textbf
  {\bibinfo {volume} {409}},\ \bibinfo {pages} {46} (\bibinfo {year}
  {2001})}\BibitemShut {NoStop}%
\bibitem [{\citenamefont {Specht}\ \emph {et~al.}(2011)\citenamefont {Specht},
  \citenamefont {N\"{o}lleke}, \citenamefont {Reiserer}, \citenamefont
  {Uphoff}, \citenamefont {Figueroa}, \citenamefont {Ritter},\ and\
  \citenamefont {Rempe}}]{Specht2011}%
  \BibitemOpen
  \bibfield  {author} {\bibinfo {author} {\bibfnamefont {H.~P.}\ \bibnamefont
  {Specht}}, \bibinfo {author} {\bibfnamefont {C.}~\bibnamefont {N\"{o}lleke}},
  \bibinfo {author} {\bibfnamefont {A.}~\bibnamefont {Reiserer}}, \bibinfo
  {author} {\bibfnamefont {M.}~\bibnamefont {Uphoff}}, \bibinfo {author}
  {\bibfnamefont {E.}~\bibnamefont {Figueroa}}, \bibinfo {author}
  {\bibfnamefont {S.}~\bibnamefont {Ritter}}, \ and\ \bibinfo {author}
  {\bibfnamefont {G.}~\bibnamefont {Rempe}},\ }\href@noop {} {\bibfield
  {journal} {\bibinfo  {journal} {Nature}\ }\textbf {\bibinfo {volume} {473}},\
  \bibinfo {pages} {190} (\bibinfo {year} {2011})}\BibitemShut {NoStop}%
\bibitem [{\citenamefont {Zhong}\ \emph {et~al.}(2015)\citenamefont {Zhong},
  \citenamefont {Hedges}, \citenamefont {Ahlefeldt}, \citenamefont
  {Bartholomew}, \citenamefont {Beavan}, \citenamefont {Wittig}, \citenamefont
  {Longdell},\ and\ \citenamefont {Sellars}}]{Zhong2015}%
  \BibitemOpen
  \bibfield  {author} {\bibinfo {author} {\bibfnamefont {M.}~\bibnamefont
  {Zhong}}, \bibinfo {author} {\bibfnamefont {M.~P.}\ \bibnamefont {Hedges}},
  \bibinfo {author} {\bibfnamefont {R.~L.}\ \bibnamefont {Ahlefeldt}}, \bibinfo
  {author} {\bibfnamefont {J.~G.}\ \bibnamefont {Bartholomew}}, \bibinfo
  {author} {\bibfnamefont {S.~E.}\ \bibnamefont {Beavan}}, \bibinfo {author}
  {\bibfnamefont {S.~M.}\ \bibnamefont {Wittig}}, \bibinfo {author}
  {\bibfnamefont {J.~J.}\ \bibnamefont {Longdell}}, \ and\ \bibinfo {author}
  {\bibfnamefont {M.~J.}\ \bibnamefont {Sellars}},\ }\href
  {http://dx.doi.org/10.1038/nature14025 10.1038/nature14025} {\bibfield
  {journal} {\bibinfo  {journal} {Nature}\ }\textbf {\bibinfo {volume} {517}},\
  \bibinfo {pages} {177} (\bibinfo {year} {2015})}\BibitemShut {NoStop}%
\bibitem [{\citenamefont {Marzban}\ \emph {et~al.}(2015)\citenamefont
  {Marzban}, \citenamefont {Bartholomew}, \citenamefont {Madden}, \citenamefont
  {Vu},\ and\ \citenamefont {Sellars}}]{Marzban2015}%
  \BibitemOpen
  \bibfield  {author} {\bibinfo {author} {\bibfnamefont {S.}~\bibnamefont
  {Marzban}}, \bibinfo {author} {\bibfnamefont {J.~G.}\ \bibnamefont
  {Bartholomew}}, \bibinfo {author} {\bibfnamefont {S.}~\bibnamefont {Madden}},
  \bibinfo {author} {\bibfnamefont {K.}~\bibnamefont {Vu}}, \ and\ \bibinfo
  {author} {\bibfnamefont {M.~J.}\ \bibnamefont {Sellars}},\ }\href {\doibase
  10.1103/PhysRevLett.115.013601} {\bibfield  {journal} {\bibinfo  {journal}
  {Phys. Rev. Lett.}\ }\textbf {\bibinfo {volume} {115}},\ \bibinfo {pages}
  {013601} (\bibinfo {year} {2015})}\BibitemShut {NoStop}%
\bibitem [{\citenamefont {Clausen}\ \emph {et~al.}(2011)\citenamefont
  {Clausen}, \citenamefont {Usmani}, \citenamefont {Bussieres}, \citenamefont
  {Sangouard}, \citenamefont {Afzelius}, \citenamefont {de~Riedmatten},\ and\
  \citenamefont {Gisin}}]{Clausen2011}%
  \BibitemOpen
  \bibfield  {author} {\bibinfo {author} {\bibfnamefont {C.}~\bibnamefont
  {Clausen}}, \bibinfo {author} {\bibfnamefont {I.}~\bibnamefont {Usmani}},
  \bibinfo {author} {\bibfnamefont {F.}~\bibnamefont {Bussieres}}, \bibinfo
  {author} {\bibfnamefont {N.}~\bibnamefont {Sangouard}}, \bibinfo {author}
  {\bibfnamefont {M.}~\bibnamefont {Afzelius}}, \bibinfo {author}
  {\bibfnamefont {H.}~\bibnamefont {de~Riedmatten}}, \ and\ \bibinfo {author}
  {\bibfnamefont {N.}~\bibnamefont {Gisin}},\ }\href@noop {} {\bibfield
  {journal} {\bibinfo  {journal} {Nature}\ }\textbf {\bibinfo {volume} {469}},\
  \bibinfo {pages} {508} (\bibinfo {year} {2011})}\BibitemShut {NoStop}%
\bibitem [{\citenamefont {Saglamyurek}\ \emph {et~al.}(2011)\citenamefont
  {Saglamyurek}, \citenamefont {Sinclair}, \citenamefont {Jin}, \citenamefont
  {Slater}, \citenamefont {Oblak}, \citenamefont {Bussi\`{e}res}, \citenamefont
  {George}, \citenamefont {Ricken}, \citenamefont {Sohler},\ and\ \citenamefont
  {Tittel}}]{Saglamyurek2011}%
  \BibitemOpen
  \bibfield  {author} {\bibinfo {author} {\bibfnamefont {E.}~\bibnamefont
  {Saglamyurek}}, \bibinfo {author} {\bibfnamefont {N.}~\bibnamefont
  {Sinclair}}, \bibinfo {author} {\bibfnamefont {J.}~\bibnamefont {Jin}},
  \bibinfo {author} {\bibfnamefont {J.~A.}\ \bibnamefont {Slater}}, \bibinfo
  {author} {\bibfnamefont {D.}~\bibnamefont {Oblak}}, \bibinfo {author}
  {\bibfnamefont {F.}~\bibnamefont {Bussi\`{e}res}}, \bibinfo {author}
  {\bibfnamefont {M.}~\bibnamefont {George}}, \bibinfo {author} {\bibfnamefont
  {R.}~\bibnamefont {Ricken}}, \bibinfo {author} {\bibfnamefont
  {W.}~\bibnamefont {Sohler}}, \ and\ \bibinfo {author} {\bibfnamefont
  {W.}~\bibnamefont {Tittel}},\ }\href@noop {} {\bibfield  {journal} {\bibinfo
  {journal} {Nature}\ }\textbf {\bibinfo {volume} {469}},\ \bibinfo {pages}
  {512} (\bibinfo {year} {2011})}\BibitemShut {NoStop}%
\bibitem [{\citenamefont {Riel\"ander}\ \emph {et~al.}(2014)\citenamefont
  {Riel\"ander}, \citenamefont {Kutluer}, \citenamefont {Ledingham},
  \citenamefont {G\"undo\ifmmode~\breve{g}\else \u{g}\fi{}an}, \citenamefont
  {Fekete}, \citenamefont {Mazzera},\ and\ \citenamefont
  {de~Riedmatten}}]{Rielander2014}%
  \BibitemOpen
  \bibfield  {author} {\bibinfo {author} {\bibfnamefont {D.}~\bibnamefont
  {Riel\"ander}}, \bibinfo {author} {\bibfnamefont {K.}~\bibnamefont
  {Kutluer}}, \bibinfo {author} {\bibfnamefont {P.~M.}\ \bibnamefont
  {Ledingham}}, \bibinfo {author} {\bibfnamefont {M.}~\bibnamefont
  {G\"undo\ifmmode~\breve{g}\else \u{g}\fi{}an}}, \bibinfo {author}
  {\bibfnamefont {J.}~\bibnamefont {Fekete}}, \bibinfo {author} {\bibfnamefont
  {M.}~\bibnamefont {Mazzera}}, \ and\ \bibinfo {author} {\bibfnamefont
  {H.}~\bibnamefont {de~Riedmatten}},\ }\href@noop {} {\bibfield  {journal}
  {\bibinfo  {journal} {Phys. Rev. Lett.}\ }\textbf {\bibinfo {volume} {112}},\
  \bibinfo {pages} {040504} (\bibinfo {year} {2014})}\BibitemShut {NoStop}%
\bibitem [{\citenamefont {Hedges}\ \emph {et~al.}(2010)\citenamefont {Hedges},
  \citenamefont {Longdell}, \citenamefont {Li},\ and\ \citenamefont
  {Sellars}}]{Hedges2010}%
  \BibitemOpen
  \bibfield  {author} {\bibinfo {author} {\bibfnamefont {M.~P.}\ \bibnamefont
  {Hedges}}, \bibinfo {author} {\bibfnamefont {J.~J.}\ \bibnamefont
  {Longdell}}, \bibinfo {author} {\bibfnamefont {Y.}~\bibnamefont {Li}}, \ and\
  \bibinfo {author} {\bibfnamefont {M.~J.}\ \bibnamefont {Sellars}},\ }\href
  {\doibase
  http://www.nature.com/nature/journal/v465/n7301/suppinfo/nature09081\_S1.html}
  {\bibfield  {journal} {\bibinfo  {journal} {Nature}\ }\textbf {\bibinfo
  {volume} {465}},\ \bibinfo {pages} {1052} (\bibinfo {year}
  {2010})}\BibitemShut {NoStop}%
\bibitem [{\citenamefont {G\"undo\ifmmode~\breve{g}\else \u{g}\fi{}an}\ \emph
  {et~al.}(2015)\citenamefont {G\"undo\ifmmode~\breve{g}\else \u{g}\fi{}an},
  \citenamefont {Ledingham}, \citenamefont {Kutluer}, \citenamefont {Mazzera},\
  and\ \citenamefont {de~Riedmatten}}]{Gundogan2015}%
  \BibitemOpen
  \bibfield  {author} {\bibinfo {author} {\bibfnamefont {M.}~\bibnamefont
  {G\"undo\ifmmode~\breve{g}\else \u{g}\fi{}an}}, \bibinfo {author}
  {\bibfnamefont {P.~M.}\ \bibnamefont {Ledingham}}, \bibinfo {author}
  {\bibfnamefont {K.}~\bibnamefont {Kutluer}}, \bibinfo {author} {\bibfnamefont
  {M.}~\bibnamefont {Mazzera}}, \ and\ \bibinfo {author} {\bibfnamefont
  {H.}~\bibnamefont {de~Riedmatten}},\ }\href {\doibase
  10.1103/PhysRevLett.114.230501} {\bibfield  {journal} {\bibinfo  {journal}
  {Phys. Rev. Lett.}\ }\textbf {\bibinfo {volume} {114}},\ \bibinfo {pages}
  {230501} (\bibinfo {year} {2015})}\BibitemShut {NoStop}%
\bibitem [{\citenamefont {Laplane}\ \emph {et~al.}(2016)\citenamefont
  {Laplane}, \citenamefont {Jobez}, \citenamefont {Etesse}, \citenamefont
  {Timoney}, \citenamefont {Gisin},\ and\ \citenamefont
  {Afzelius}}]{Afzelius2016}%
  \BibitemOpen
  \bibfield  {author} {\bibinfo {author} {\bibfnamefont {C.}~\bibnamefont
  {Laplane}}, \bibinfo {author} {\bibfnamefont {P.}~\bibnamefont {Jobez}},
  \bibinfo {author} {\bibfnamefont {J.}~\bibnamefont {Etesse}}, \bibinfo
  {author} {\bibfnamefont {N.}~\bibnamefont {Timoney}}, \bibinfo {author}
  {\bibfnamefont {N.}~\bibnamefont {Gisin}}, \ and\ \bibinfo {author}
  {\bibfnamefont {M.}~\bibnamefont {Afzelius}},\ }\href@noop {} {\bibfield
  {journal} {\bibinfo  {journal} {New Journal of Physics}\ }\textbf {\bibinfo
  {volume} {18}},\ \bibinfo {pages} {013006} (\bibinfo {year}
  {2016})}\BibitemShut {NoStop}%
\bibitem [{\citenamefont {Duan}\ \emph {et~al.}(2001)\citenamefont {Duan},
  \citenamefont {Lukin}, \citenamefont {Cirac},\ and\ \citenamefont
  {Zoller}}]{Duan2001}%
  \BibitemOpen
  \bibfield  {author} {\bibinfo {author} {\bibfnamefont {L.~M.}\ \bibnamefont
  {Duan}}, \bibinfo {author} {\bibfnamefont {M.~D.}\ \bibnamefont {Lukin}},
  \bibinfo {author} {\bibfnamefont {J.~I.}\ \bibnamefont {Cirac}}, \ and\
  \bibinfo {author} {\bibfnamefont {P.}~\bibnamefont {Zoller}},\ }\href
  {\doibase
  http://www.nature.com/nature/journal/v414/n6862/suppinfo/414413a0\_S1.html}
  {\bibfield  {journal} {\bibinfo  {journal} {Nature}\ }\textbf {\bibinfo
  {volume} {414}},\ \bibinfo {pages} {413} (\bibinfo {year}
  {2001})}\BibitemShut {NoStop}%
\bibitem [{\citenamefont {Goldschmidt}\ \emph {et~al.}(2013)\citenamefont
  {Goldschmidt}, \citenamefont {Beavan}, \citenamefont {Polyakov},
  \citenamefont {Migdall},\ and\ \citenamefont {Sellars}}]{Goldschmidt2013}%
  \BibitemOpen
  \bibfield  {author} {\bibinfo {author} {\bibfnamefont {E.~A.}\ \bibnamefont
  {Goldschmidt}}, \bibinfo {author} {\bibfnamefont {S.~E.}\ \bibnamefont
  {Beavan}}, \bibinfo {author} {\bibfnamefont {S.~V.}\ \bibnamefont
  {Polyakov}}, \bibinfo {author} {\bibfnamefont {A.~L.}\ \bibnamefont
  {Migdall}}, \ and\ \bibinfo {author} {\bibfnamefont {M.~J.}\ \bibnamefont
  {Sellars}},\ }\href@noop {} {\bibfield  {journal} {\bibinfo  {journal} {Opt.
  Express}\ }\textbf {\bibinfo {volume} {21}},\ \bibinfo {pages} {10087}
  (\bibinfo {year} {2013})}\BibitemShut {NoStop}%
\bibitem [{\citenamefont {Ledingham}\ \emph {et~al.}(2010)\citenamefont
  {Ledingham}, \citenamefont {Naylor}, \citenamefont {Longdell}, \citenamefont
  {Beavan},\ and\ \citenamefont {Sellars}}]{Ledingham2010}%
  \BibitemOpen
  \bibfield  {author} {\bibinfo {author} {\bibfnamefont {P.~M.}\ \bibnamefont
  {Ledingham}}, \bibinfo {author} {\bibfnamefont {W.~R.}\ \bibnamefont
  {Naylor}}, \bibinfo {author} {\bibfnamefont {J.~J.}\ \bibnamefont
  {Longdell}}, \bibinfo {author} {\bibfnamefont {S.~E.}\ \bibnamefont
  {Beavan}}, \ and\ \bibinfo {author} {\bibfnamefont {M.~J.}\ \bibnamefont
  {Sellars}},\ }\href {\doibase 10.1103/PhysRevA.81.012301} {\bibfield
  {journal} {\bibinfo  {journal} {Phys. Rev. A}\ }\textbf {\bibinfo {volume}
  {81}},\ \bibinfo {pages} {012301} (\bibinfo {year} {2010})}\BibitemShut
  {NoStop}%
\bibitem [{\citenamefont {Beavan}\ \emph {et~al.}(2012)\citenamefont {Beavan},
  \citenamefont {Hedges},\ and\ \citenamefont {Sellars}}]{Beavan2012}%
  \BibitemOpen
  \bibfield  {author} {\bibinfo {author} {\bibfnamefont {S.~E.}\ \bibnamefont
  {Beavan}}, \bibinfo {author} {\bibfnamefont {M.~P.}\ \bibnamefont {Hedges}},
  \ and\ \bibinfo {author} {\bibfnamefont {M.~J.}\ \bibnamefont {Sellars}},\
  }\href {\doibase 10.1103/PhysRevLett.109.093603} {\bibfield  {journal}
  {\bibinfo  {journal} {Phys. Rev. Lett.}\ }\textbf {\bibinfo {volume} {109}},\
  \bibinfo {pages} {093603} (\bibinfo {year} {2012})}\BibitemShut {NoStop}%
\bibitem [{\citenamefont {Kurnit}\ \emph {et~al.}(1964)\citenamefont {Kurnit},
  \citenamefont {Abella},\ and\ \citenamefont {Hartmann}}]{Kurnit1964}%
  \BibitemOpen
  \bibfield  {author} {\bibinfo {author} {\bibfnamefont {N.~A.}\ \bibnamefont
  {Kurnit}}, \bibinfo {author} {\bibfnamefont {I.~D.}\ \bibnamefont {Abella}},
  \ and\ \bibinfo {author} {\bibfnamefont {S.~R.}\ \bibnamefont {Hartmann}},\
  }\href {\doibase 10.1103/PhysRevLett.13.567} {\bibfield  {journal} {\bibinfo
  {journal} {Phys. Rev. Lett.}\ }\textbf {\bibinfo {volume} {13}},\ \bibinfo
  {pages} {567} (\bibinfo {year} {1964})}\BibitemShut {NoStop}%
\bibitem [{\citenamefont {Ledingham}\ \emph {et~al.}(2012)\citenamefont
  {Ledingham}, \citenamefont {Naylor},\ and\ \citenamefont
  {Longdell}}]{Ledingham2012}%
  \BibitemOpen
  \bibfield  {author} {\bibinfo {author} {\bibfnamefont {P.~M.}\ \bibnamefont
  {Ledingham}}, \bibinfo {author} {\bibfnamefont {W.~R.}\ \bibnamefont
  {Naylor}}, \ and\ \bibinfo {author} {\bibfnamefont {J.~J.}\ \bibnamefont
  {Longdell}},\ }\href {\doibase 10.1103/PhysRevLett.109.093602} {\bibfield
  {journal} {\bibinfo  {journal} {Phys. Rev. Lett.}\ }\textbf {\bibinfo
  {volume} {109}},\ \bibinfo {pages} {093602} (\bibinfo {year}
  {2012})}\BibitemShut {NoStop}%
\bibitem [{\citenamefont {Equall}\ \emph {et~al.}(1995)\citenamefont {Equall},
  \citenamefont {Cone},\ and\ \citenamefont {Macfarlane}}]{Equall1995}%
  \BibitemOpen
  \bibfield  {author} {\bibinfo {author} {\bibfnamefont {R.~W.}\ \bibnamefont
  {Equall}}, \bibinfo {author} {\bibfnamefont {R.~L.}\ \bibnamefont {Cone}}, \
  and\ \bibinfo {author} {\bibfnamefont {R.~M.}\ \bibnamefont {Macfarlane}},\
  }\href {http://link.aps.org/doi/10.1103/PhysRevB.52.3963} {\bibfield
  {journal} {\bibinfo  {journal} {Phys. Rev. B}\ }\textbf {\bibinfo {volume}
  {52}},\ \bibinfo {pages} {3963} (\bibinfo {year} {1995})}\BibitemShut
  {NoStop}%
\bibitem [{\citenamefont {Holliday}\ \emph {et~al.}(1993)\citenamefont
  {Holliday}, \citenamefont {Croci}, \citenamefont {Vauthey},\ and\
  \citenamefont {Wild}}]{Holliday1993}%
  \BibitemOpen
  \bibfield  {author} {\bibinfo {author} {\bibfnamefont {K.}~\bibnamefont
  {Holliday}}, \bibinfo {author} {\bibfnamefont {M.}~\bibnamefont {Croci}},
  \bibinfo {author} {\bibfnamefont {E.}~\bibnamefont {Vauthey}}, \ and\
  \bibinfo {author} {\bibfnamefont {U.~P.}\ \bibnamefont {Wild}},\ }\href
  {\doibase 10.1103/PhysRevB.47.14741} {\bibfield  {journal} {\bibinfo
  {journal} {Phys. Rev. B}\ }\textbf {\bibinfo {volume} {47}},\ \bibinfo
  {pages} {14741} (\bibinfo {year} {1993})}\BibitemShut {NoStop}%
\bibitem [{\citenamefont {Fraval}(2006)}]{FravalE}%
  \BibitemOpen
  \bibfield  {author} {\bibinfo {author} {\bibfnamefont {E.}~\bibnamefont
  {Fraval}},\ }\emph {\bibinfo {title} {Minimising the Decoherence of Rare
  Earth Ion Solid State Spin Qubits}},\ \href@noop {} {Ph.D. thesis},\ \bibinfo
   {school} {Australian National University} (\bibinfo {year}
  {2006})\BibitemShut {NoStop}%
\bibitem [{\citenamefont {Nilsson}\ \emph {et~al.}(2004)\citenamefont
  {Nilsson}, \citenamefont {Rippe}, \citenamefont {Kr\"oll}, \citenamefont
  {Klieber},\ and\ \citenamefont {Suter}}]{Nilsson2004}%
  \BibitemOpen
  \bibfield  {author} {\bibinfo {author} {\bibfnamefont {M.}~\bibnamefont
  {Nilsson}}, \bibinfo {author} {\bibfnamefont {L.}~\bibnamefont {Rippe}},
  \bibinfo {author} {\bibfnamefont {S.}~\bibnamefont {Kr\"oll}}, \bibinfo
  {author} {\bibfnamefont {R.}~\bibnamefont {Klieber}}, \ and\ \bibinfo
  {author} {\bibfnamefont {D.}~\bibnamefont {Suter}},\ }\href {\doibase
  10.1103/PhysRevB.70.214116} {\bibfield  {journal} {\bibinfo  {journal} {Phys.
  Rev. B}\ }\textbf {\bibinfo {volume} {70}},\ \bibinfo {pages} {214116}
  (\bibinfo {year} {2004})}\BibitemShut {NoStop}%
\bibitem [{\citenamefont {Beavan}\ \emph {et~al.}(2011)\citenamefont {Beavan},
  \citenamefont {Ledingham}, \citenamefont {Longdell},\ and\ \citenamefont
  {Sellars}}]{Beavan2011}%
  \BibitemOpen
  \bibfield  {author} {\bibinfo {author} {\bibfnamefont {S.~E.}\ \bibnamefont
  {Beavan}}, \bibinfo {author} {\bibfnamefont {P.~M.}\ \bibnamefont
  {Ledingham}}, \bibinfo {author} {\bibfnamefont {J.~J.}\ \bibnamefont
  {Longdell}}, \ and\ \bibinfo {author} {\bibfnamefont {M.~J.}\ \bibnamefont
  {Sellars}},\ }\href {\doibase 10.1364/OL.36.001272} {\bibfield  {journal}
  {\bibinfo  {journal} {Opt. Lett.}\ }\textbf {\bibinfo {volume} {36}},\
  \bibinfo {pages} {1272} (\bibinfo {year} {2011})}\BibitemShut {NoStop}%
\bibitem [{Note1()}]{Note1}%
  \BibitemOpen
  \bibinfo {note} {See Supplementary Material at URL for details of gain
  measurements, phase correction, the windowing of the ASE and RASE signals and
  the inseparability criterion model.}\BibitemShut {Stop}%
\bibitem [{\citenamefont {Duan}\ \emph {et~al.}(2000)\citenamefont {Duan},
  \citenamefont {Giedke}, \citenamefont {Cirac},\ and\ \citenamefont
  {Zoller}}]{Duan2000}%
  \BibitemOpen
  \bibfield  {author} {\bibinfo {author} {\bibfnamefont {L.-M.}\ \bibnamefont
  {Duan}}, \bibinfo {author} {\bibfnamefont {G.}~\bibnamefont {Giedke}},
  \bibinfo {author} {\bibfnamefont {J.~I.}\ \bibnamefont {Cirac}}, \ and\
  \bibinfo {author} {\bibfnamefont {P.}~\bibnamefont {Zoller}},\ }\href
  {\doibase 10.1103/PhysRevLett.84.2722} {\bibfield  {journal} {\bibinfo
  {journal} {Phys. Rev. Lett.}\ }\textbf {\bibinfo {volume} {84}},\ \bibinfo
  {pages} {2722} (\bibinfo {year} {2000})}\BibitemShut {NoStop}%
\bibitem [{\citenamefont {Yuen}\ and\ \citenamefont {Chan}(1983)}]{Yuen1983}%
  \BibitemOpen
  \bibfield  {author} {\bibinfo {author} {\bibfnamefont {H.~P.}\ \bibnamefont
  {Yuen}}\ and\ \bibinfo {author} {\bibfnamefont {V.~W.~S.}\ \bibnamefont
  {Chan}},\ }\href@noop {} {\bibfield  {journal} {\bibinfo  {journal} {Opt.
  Lett.}\ }\textbf {\bibinfo {volume} {8}},\ \bibinfo {pages} {177} (\bibinfo
  {year} {1983})}\BibitemShut {NoStop}%
\bibitem [{\citenamefont {Stevenson}\ \emph {et~al.}(2014)\citenamefont
  {Stevenson}, \citenamefont {Hush}, \citenamefont {Carvalho}, \citenamefont
  {Beavan}, \citenamefont {Sellars},\ and\ \citenamefont
  {Hope}}]{Stevenson2014}%
  \BibitemOpen
  \bibfield  {author} {\bibinfo {author} {\bibfnamefont {R.~N.}\ \bibnamefont
  {Stevenson}}, \bibinfo {author} {\bibfnamefont {M.~R.}\ \bibnamefont {Hush}},
  \bibinfo {author} {\bibfnamefont {a.~R.~R.}\ \bibnamefont {Carvalho}},
  \bibinfo {author} {\bibfnamefont {S.~E.}\ \bibnamefont {Beavan}}, \bibinfo
  {author} {\bibfnamefont {M.~J.}\ \bibnamefont {Sellars}}, \ and\ \bibinfo
  {author} {\bibfnamefont {J.~J.}\ \bibnamefont {Hope}},\ }\href {\doibase
  10.1088/1367-2630/16/3/033042} {\bibfield  {journal} {\bibinfo  {journal}
  {New Journal of Physics}\ }\textbf {\bibinfo {volume} {16}},\ \bibinfo
  {pages} {033042} (\bibinfo {year} {2014})}\BibitemShut {NoStop}%
\bibitem [{\citenamefont {Williamson}\ and\ \citenamefont
  {Longdell}(2014)}]{Williamson2014}%
  \BibitemOpen
  \bibfield  {author} {\bibinfo {author} {\bibfnamefont {L.~A.}\ \bibnamefont
  {Williamson}}\ and\ \bibinfo {author} {\bibfnamefont {J.~J.}\ \bibnamefont
  {Longdell}},\ }\href {http://arxiv.org/abs/1403.6872} {\bibfield  {journal}
  {\bibinfo  {journal} {New J. Phys.}\ }\textbf {\bibinfo {volume} {16}},\
  \bibinfo {pages} {073046} (\bibinfo {year} {2014})}\BibitemShut {NoStop}%
\bibitem [{\citenamefont {Bao}\ \emph {et~al.}(2012)\citenamefont {Bao},
  \citenamefont {Reingruber}, \citenamefont {Dietrich}, \citenamefont {Rui},
  \citenamefont {D\"{u}ck}, \citenamefont {Strassel}, \citenamefont {Li},
  \citenamefont {Liu}, \citenamefont {Zhao},\ and\ \citenamefont
  {Pan}}]{Bao2012}%
  \BibitemOpen
  \bibfield  {author} {\bibinfo {author} {\bibfnamefont {X.-H.}\ \bibnamefont
  {Bao}}, \bibinfo {author} {\bibfnamefont {A.}~\bibnamefont {Reingruber}},
  \bibinfo {author} {\bibfnamefont {P.}~\bibnamefont {Dietrich}}, \bibinfo
  {author} {\bibfnamefont {J.}~\bibnamefont {Rui}}, \bibinfo {author}
  {\bibfnamefont {A.}~\bibnamefont {D\"{u}ck}}, \bibinfo {author}
  {\bibfnamefont {T.}~\bibnamefont {Strassel}}, \bibinfo {author}
  {\bibfnamefont {L.}~\bibnamefont {Li}}, \bibinfo {author} {\bibfnamefont
  {N.-L.}\ \bibnamefont {Liu}}, \bibinfo {author} {\bibfnamefont
  {B.}~\bibnamefont {Zhao}}, \ and\ \bibinfo {author} {\bibfnamefont {J.-W.}\
  \bibnamefont {Pan}},\ }\href {\doibase 10.1038/nphys2324} {\bibfield
  {journal} {\bibinfo  {journal} {Nature Physics}\ }\textbf {\bibinfo {volume}
  {8}},\ \bibinfo {pages} {517} (\bibinfo {year} {2012})}\BibitemShut {NoStop}%
\bibitem [{\citenamefont {Fraval}\ \emph {et~al.}(2004)\citenamefont {Fraval},
  \citenamefont {Sellars},\ and\ \citenamefont {Longdell}}]{Fraval2004}%
  \BibitemOpen
  \bibfield  {author} {\bibinfo {author} {\bibfnamefont {E.}~\bibnamefont
  {Fraval}}, \bibinfo {author} {\bibfnamefont {M.~J.}\ \bibnamefont {Sellars}},
  \ and\ \bibinfo {author} {\bibfnamefont {J.~J.}\ \bibnamefont {Longdell}},\
  }\href {\doibase 10.1103/PhysRevLett.92.077601} {\bibfield  {journal}
  {\bibinfo  {journal} {Phys. Rev. Lett.}\ }\textbf {\bibinfo {volume} {92}},\
  \bibinfo {pages} {077601} (\bibinfo {year} {2004})}\BibitemShut {NoStop}%
\bibitem [{\citenamefont {Simon}\ \emph {et~al.}(2007)\citenamefont {Simon},
  \citenamefont {de~Riedmatten}, \citenamefont {Afzelius}, \citenamefont
  {Sangouard}, \citenamefont {Zbinden},\ and\ \citenamefont
  {Gisin}}]{Simon2007}%
  \BibitemOpen
  \bibfield  {author} {\bibinfo {author} {\bibfnamefont {C.}~\bibnamefont
  {Simon}}, \bibinfo {author} {\bibfnamefont {H.}~\bibnamefont
  {de~Riedmatten}}, \bibinfo {author} {\bibfnamefont {M.}~\bibnamefont
  {Afzelius}}, \bibinfo {author} {\bibfnamefont {N.}~\bibnamefont {Sangouard}},
  \bibinfo {author} {\bibfnamefont {H.}~\bibnamefont {Zbinden}}, \ and\
  \bibinfo {author} {\bibfnamefont {N.}~\bibnamefont {Gisin}},\ }\href@noop {}
  {\bibfield  {journal} {\bibinfo  {journal} {Phys. Rev. Lett.}\ }\textbf
  {\bibinfo {volume} {98}},\ \bibinfo {pages} {190503} (\bibinfo {year}
  {2007})}\BibitemShut {NoStop}%
\end{thebibliography}%

\end{document}